\documentclass{pasj00}

\SetRunningHead{S. Minami et al.}{Suzaku observation of G355.6$-$0.0}
\Received{yyyy/XX/XX}%{yyyy/mm/dd}
\Accepted{yyyy/XX/XX}%{yyyy/mm/dd}
\begin{document}
\title{X-Ray Observation of the Galactic Supernova Remnant G355.6$-$0.0 with Suzaku}

%%% begin:list of authors
% Do NOT capitalize all letters in "textsc".
\author{Sari \textsc{Minami}, Naomi \textsc{Ota}, and Shigeo \textsc{Yamauchi}}
\affil{Department of Physics, Faculty of Science, Nara Women's University, Kitauoyanishi-machi, Nara, Nara 630-8506}
\email{nas\_minami@cc.nara-wu.ac.jp}
\and
\author{Katsuji \textsc{Koyama}}
\affil{Department of Physics, Graduate School of Science, Kyoto University, 
Kitashirakawa-Oiwake-cho, Sakyo-ku, Kyoto 606-8502\\
Department of Earth and Space Science, Graduate School of Science, Osaka University, \\
1-1 Machikaneyama-cho, Toyonaka, Osaka 560-0043}

%% `\KeyWords{}' always has to be placed before `\maketitle'.
 %Do NOT move this preamble from here!
\maketitle

\begin{abstract}
We present results of the Galactic supernova remnant (SNR) G355.6$-$0.0 observed with Suzaku.
We resolved the diffuse emission detected with ASCA into two objects, G355.6$-$0.0 and a point-like source, Suzaku J173530$-$3236.
The X-ray emission from G355.6$-$0.0 exhibits a center-filled morphology within the radio shell.
The X-ray spectrum is well represented by a thin thermal plasma model with enhanced metal abundances.
The spatial and spectral properties imply that G355.6$-$0.0 is a member of the mixed-morphology SNRs.
The $N_{\rm H}$ value of $\sim$6$\times$10$^{22}$ cm$^{-2}$ supports that G355.6$-$0.0 is a distant SNR.
Suzaku J173530$-$3236 exhibits a hard X-ray spectrum with a strong Fe emission line, similar to 
those of cataclysmic variables.
The $N_{\rm H}$ value of $\sim$3$\times$10$^{22}$ cm$^{-2}$ is smaller than that of G355.6$-$0.0, and hence Suzaku J173530$-$3236 is located at the near side of G355.6$-$0.0.
\end{abstract}

%%%%%%%%%%%%%%%%%%%%%%%%%%%%%%%%%%%%%%%%%%%%%%%%%%	
\section{Introduction}

Supernovae (SNe) and supernova remnants (SNRs) are important objects, 
because they are suppliers of a large amount of energies into interstellar space and 
the main sites of heavy element production and acceleration of high-energy particles.	
SNRs are typically classified into two categories based on their structures and spectral features.
One is called Crab-like (Plerionic) SNRs that have a pulsar at the heart of the object. The X-ray and radio are both non-thermal emissions with center-filled morphologies.
The other class has a shell-like structure in both X-ray and radio bands, and hence is called shell-like SNRs.
The X-ray emission is thin thermal emission, originating from shock-heated plasma, while the radio is synchrotron emission from accelerated particles.
\citet{mm} proposed the other class of SNRs, which has a centrally peaked thermal X-rays in a non-thermal radio shell, 
and hence called Mixed-Morphology (MM) SNRs.
A surface brightness distribution of MM SNRs is constant or increases toward the remnant center, contrary to the standard Sedov model.
Several models for the origin of the MM SNRs have been proposed \citep{MM1, MM2, MM3}, but formation mechanisms and  evolutional process are debatable.  

In order to investigate the properties and address a unified scenario of the MM- SNRs, systematic X-ray studies, particularly for faint and old SNRs, would be essential.  
However, these studies using the previous X-ray satellite have been very limited, because most of the SNRs are located near or on the Galactic plane, where the Galactic X-ray background is not negligible. 
Also candidate SNRs are all soft X-ray sources, and the soft X-ray bands are strongly absorbed by interstellar medium on the plane. 

We started systematic survey observation for study of various MM-SNRs on the Galactic plane using the Suzaku satellite. The Suzaku satellite has high sensitivity for detecting X-ray emissions, a better energy resolution, and lower and more stable detector background than the previous X-ray astronomy satellites \citep{mitsuda}, and hence would be  the most suitable facility for observations of extended and/or faint MM-SNRs. 

G355.6$-$0.0 was first discovered by the Molonglo Observatory Synthesis Telescope (MOST) which made an 843-MHz survey of the Galactic Center region \citep{gray}.
The MOST image indicates a shell-like structure with a size of $6'\times8'$ and the spectral feature supports that G355.6$-$0.0 is an SNR.
X-ray emission at the position of G355.6$-$0.0 was first detected in the ASCA Galactic plane survey \citep{asca1}.
\citet{asca2} reported that the spectrum was represented by a two-temperature thin thermal plasma model with temperatures of $0.7^{+0.3}_{-0.4}$ keV and 7.5 ($>$ 1.7) keV and solar abundances.
Due to limited photon statistics, however, the origin of X-ray emission is still unknown. 
In this paper, we report the Suzaku results of G355.6$-$0.0.
All of the errors are at the 90\% confidence level throughout this paper.
%, unless otherwise mentioned.

%image
\begin{figure*}
   \begin{center}
      \FigureFile(160mm,160mm){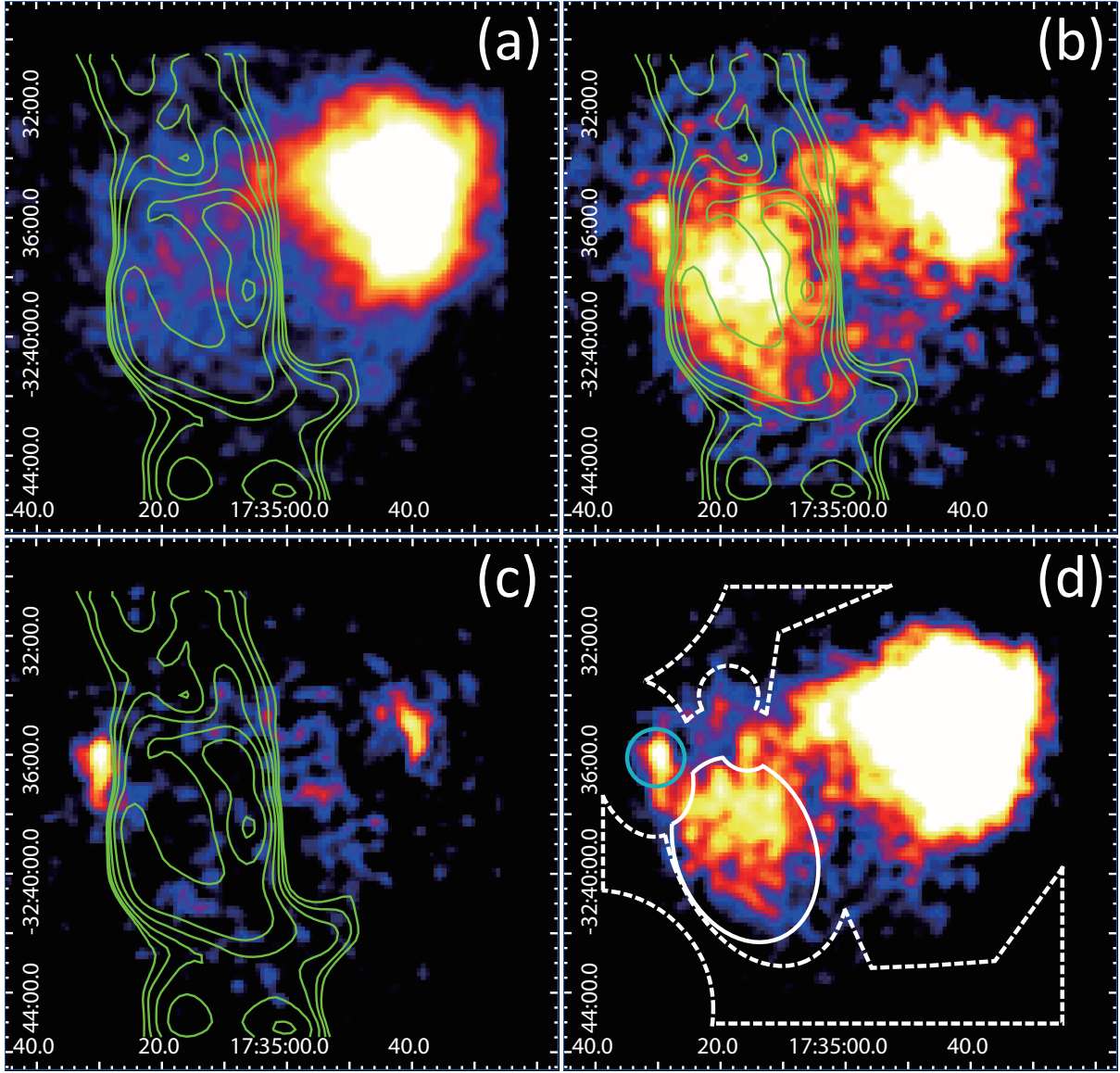}
   \end{center}
   \caption{Suzaku XIS image of G355.6$-$0.0 in the 0.7--1.0 (a), 1.0--5.0 (b), 5.0--10.0
 (c), and 0.7--10.0 (d) keV energy bands. The images were smoothed with a Gaussian distribution of $\sigma$ = 24$''$. 
 The color scale is logarithmic, while the coordinates are J2000.0. Neither the background subtraction nor the vignetting correction were made. The contours in (a)--(c) show the radio band image at 1.4GHz obtained from the NRAO VLA Sky Survey (NVSS) \citep{radio}. The white and the blue solid lines in (d) show the regions from which the X-ray spectra of G355.6$-$0.0 and Suzaku J173530$-$3236 were extracted, respectively, while  
the  dashed line shows the background region. } 
\label{image}
\end{figure*}

%%%%%%%%%%%%%%%%%%%%%%%%%%%%%%%%%%%%%%%%%%%%%%%%%%	
\section{Observation and Data Reduction}

Suzaku observed G355.6$-$0.0 with the X-ray Imaging Spectrometers (XIS: \cite{koyama}) at the focal planes of the X-ray Telescope (XRT: \cite{XRT}) from 2010 February 19 12:33:00 to 2010 February 20 15:06:00 (observation ID 504098010).
The XIS has a high energy resolution over the 0.2--10 keV energy range with a field of view (FOV) of $17.8'\times17.8'$.
XIS1 is a back-side illuminated (BI) CCD chip, while XIS0, 2, and 3 are front-side illuminated (FI) CCDs.
We used XIS0, 1, 3 because XIS2 has been out of function since 2006 November. Also a fraction of the XIS0 region has not been available since 2009 June 23 due to a damage by possible impact of micro meteorite.
The XIS was operated in the normal full-frame clocking mode with a spaced-row charge injection technique \citep{SCI}.
	
Data reduction was performed using the HEAsoft version 6.11 and the processed data version 2.4.
We reprocessed the data using {\tt xispi} software version 2009-02-28 and the calibration database released on 2011 November 9.
We rejected the data taken at the South Atlantic Anomaly, during the Earth occultation, and at the low elevation angle from the Earth rim of $<$5$^{\circ}$ (night Earth) and $<$20$^{\circ}$ (day Earth).
After removing hot and flickering pixels of the CCD, we selected grades 0, 2, 3, 4, and 6 events as the X-ray signals.
The exposure time after these screenings was 52.5 ksec for each XIS detector. 

%%%%%%%%%%%%%%%%%%%%%%%%%%%%%%%%%%%%%%%%%%%%%%%%%%	
\section{Results}

\subsection{X-Ray Image}

Figure 1 shows the XIS images of G355.6$-$0.0 in the 0.7--1.0, 1.0--5.0, 5.0--10.0, and 0.7--10.0 keV energy bands.
For maximizing photon statistics, the data of XIS0, 1, and 3 were combined.
The radio band image at 1.4 GHz \citep{radio} is shown by the contours.
The brightest source located at the northwest of the FOV is an open cluster NGC 6383 whose X-rays are dominant in the soft X-ray band.
X-ray emission from G355.6$-$0.0 is clearly seen in the 1--5 keV band
with a center-filled morphology of an apparent size of $\sim4'\times6'$ within the radio shell, while no X-ray emission was found, neither  below 1 keV nor above 5 keV.   
Near the northeast of the SNR, a point-like source was found at (RA, Dec)$_{\rm J2000.0}$ = ($\timeform{17h35m30.3s}$, $\timeform{-32D36'03"}$) 
(typical uncertainty of $\sim$19$''$, \cite{uchiyama2008})
, and is named Suzaku J173530$-$3236.
We searched for a counterpart using the NED and the SIMBAD data system, and found XMM J173530.9$-$323558 \citep{xmm} 
within an error circle.
% of 8.6 $''$. %<--'¾"cæ¶ver
%within the error region (separation angle = 8.6$''$). <--ŽR"àæ¶ver

\subsection{G355.6$-$0.0}

%[table1] 
\begin{table*}
\begin{center}
\caption{Results of a model fit with a thermal bremsstrahlung and Gaussian line model and line identification.}\label{tab:line}
  \begin{tabular}{lccc} \hline
  {Parameter} & \multicolumn{2}{c}{Value} & {Line identification} \\ \hline
                         & G355-a                             & G355-b & \\ \hline
   $N_{\rm H}$\ (\(\times\)10\( ^{22} \)cm\( ^{-2}\)) & $4.8^{+1.4}_{-1.2}$ & $4.9^{+1.1}_{-1.1}$  &  \\ 
   {\it kT}\ (keV) & $0.76^{+0.27}_{-0.18}$ & $0.73^{+0.20}_{-0.15}$  & \\
   $E_{\rm line1}$\ (keV) & $1.85^{+0.01}_{-0.01}$ & $1.85^{+0.01}_{-0.01}$  & Si\ {\footnotesize XIII}\ K\(\alpha\) \\ 
   $E_{\rm line2}$\ (keV) & $2.19^{+0.04}_{-0.04}$ & $2.20^{+0.05}_{-0.04}$  & Si\ {\footnotesize XIII}\ K\(\beta\) \\ 
   $E_{\rm line3}$\ (keV) & $2.46^{+0.01}_{-0.02}$ & $2.46^{+0.01}_{-0.02}$  & S\ {\footnotesize XV}\ K\(\alpha\) \\ 
   $E_{\rm line4}$\ (keV) & 2.87$^{\ast}$ & $2.91^{+0.10}_{-0.08}$  & S\ {\footnotesize XV}\ K\(\beta\) \\ 
   $E_{\rm line5}$\ (keV) & $3.11^{+0.03}_{-0.03}$ & $3.11^{+0.02}_{-0.03}$  & Ar\ {\footnotesize XVII}\ K\(\alpha\) \\ 
   $E_{\rm line6}$\ (keV) & $3.90^{+0.05}_{-0.05}$ & $3.91^{+0.04}_{-0.03}$  & Ca\ {\footnotesize XIX}\ K\(\alpha\) \\ 
   \(\chi^2\) /d.o.f. & 64.1/71 & 89.3/79 & \\ \hline
   \multicolumn{3}{@{}l@{}}{\hbox to 0pt{\parbox{85mm}{\footnotesize
	    \footnotemark[$*$] Fixed to $1.17 \times E_{\rm line3}$.
	    \par\noindent
  }\hss}}
  \end{tabular}
 \end{center}
 \end{table*}
 
\begin{figure}
   \begin{center}
      \FigureFile(80mm,80mm){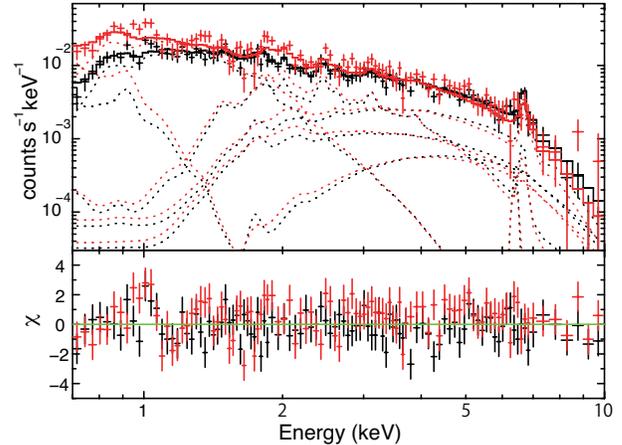}
   \end{center}
   \caption{NXB-subtracted near-sky background spectra and the best-fit model (upper), and 
   residuals from the best-fit model (lower). 
   Data of XIS0+3 and XIS1 are indicated in black and red, respectively.}\label{}
\end{figure}

%spectrum
\begin{figure*}
   \begin{center}
      \FigureFile(160mm,80mm){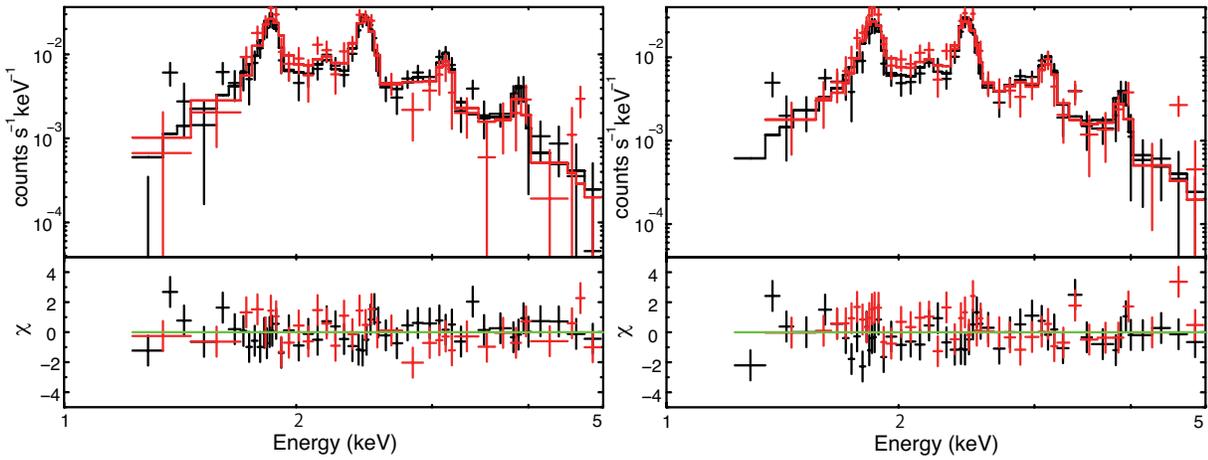}
%      \FigureFile(80mm,80mm){ronbun_vapec_b525.eps}
   \end{center}
   \caption{XIS spectra of G355.6$-$0.0 and the best-fit CIE model (see table 2) (upper), and 
   residuals from the best-fit model (lower).  
   The left panel is the results of G355-a (the near-sky-background-subtracted spectra), 
   while the right panel is those of G355-b (the simulated-background-subtracted spectra).
   Data of XIS0+3 and XIS1 are indicated in black and red, respectively.}\label{}
\end{figure*}

The X-ray spectrum of G355.6$-$0.0 was extracted from an elliptical region with a major axis of 6$'$ and a minor axis of 4$'$, but X-ray photons in the regions near to Suzaku J173530$-$3236 
(=XMM J173530.9$-$323558)
 and XMM J173517.4$-$323549 \citep{xmm} were excluded (see figure 1d).
The background spectrum was extracted from a nearby source-free region 
in the same FOV (the dashed lines in figure 1d).

Response files, Redistribution Matrix Files (RMFs) and Ancillary Response Files (ARFs), were 
created using {\tt xisrmfgen} and {\tt xissimarfgen} in HEAsoft  \citep{rmfarf}, respectively.
%In making response files, we used observed image of the source region as the photon distribution data.
Since the number of X-ray counts is small, we combined the spectra obtained with XIS0 and XIS3 
and fitted the XIS0$+$3 and XIS1 spectra with a spectral model simultaneously.

Since G355.6$-$0.0 is a faint diffuse source on the Galactic plane, 
%the background should be subtracted carefully.
careful assessment of background is needed.
We constructed the non-X-ray background (NXB) for the source and the background spectra 
from the night earth data using {\tt xisnxbgen} in HEAsoft \citep{nxb}.
After subtracting the NXB, 
we corrected the difference of vignetting effects between the source and background regions 
by the method described in \citet{hyodo}.
The NXB-subtracted near-sky background spectra were subtracted from the NXB-subtracted source spectra (hereafter G355-a).

As seen in figure 1d, the background region is small 
because the bright X-ray source NGC 6383 is located in the FOV.
Hence, the photon statistics of the near-sky background spectra were limited.
In order to obtain statistically better spectra, we examined another set of background data.
\citet{uchiyama2013} modeled the blank sky spectra on the Galactic plane, consisting of the Galactic Ridge X-ray Emission (GRXE) and the cosmic X-ray background (CXB), with several components.
We fitted the NXB-subtracted near-sky background spectra with the same model as \citet{uchiyama2013}:

\medskip

\{TP ($kT$=0.09 keV)$+$TP ($kT$=0.59 keV)\} $\times$ABS1 $+$ 
\{TP ($kT$=1.33 keV)$+$TP ($kT$=6.64 keV) $+$reflected component$+$CXB$\times$ABS2\} $\times$ABS2,
 
\medskip
 
 \noindent
where TP and ABS show the thin thermal plasma model ({\tt vapec} in XSPEC), 
and photoelectric absorption ({\tt phabs} in XSPEC, \cite{phabs}), respectively.
The reflected component consisted of a power-law model and K$\alpha$ (6.4 keV) 
and K$\beta$ (7.05 keV)  lines from Fe atoms.
Free parameters are normalizations of the TP and the power-law models and the $N_{\rm H}$ value of ABS2, while 
the metal abundances, temperatures, and the $N_{\rm H}$ value of ABS1 were fixed to those in \citet{uchiyama2013}.
The normalization and the photon index of the CXB were assumed to be 10 photons $\rm{s^{-1}}$ $\rm{cm^{-2}}$ $\rm{keV^{-1}}$ $\rm{str^{-1}}$ at 1 keV and 1.4, respectively (e.g., \cite{rev}; \cite{marshall}).
The NXB-subtracted background spectra and the best-fit model are shown in figure 2.
The model well represented the spectra above 1.1 keV 
with a $\chi^2$ value of 252.66 for d.o.f. of 206.
Using the best-fit parameters, we simulated the background spectra for the source region %with exposure time of 1 Ms
and subtracted them from the NXB-subtracted source spectra (hereafter  G355-b).

As we noted, the soft X-rays from G355.6$-$0.0 are significantly absorbed
and few X-ray photons were detected 
in the $>$5.00 keV band (see figure 1). 
Therefore, we used the data in the 1.15--5.00 keV energy band.
Figure 3 shows the background-subtracted spectra.
We see many emission lines in the spectra.
To identify the emission lines, we fitted the spectra with a model consisting of bremsstrahlung and
 Gaussian lines modified by interstellar absorption.
All the intrinsic line widths were fixed to null.
The results of the spectral fit are listed in table \ref{tab:line}.
Taking account of the uncertainty of the energy calibration of the XIS and statistical errors, 
the center energies of these Gaussian lines are in agreement with the
theoretical values of the K-shell lines from He-like Si, S, Ar, and Ca \citep{lines}.  
This is the first detection of emission lines from highly ionized atoms of this SNR. 
Thus, the X-ray emission can be attributed to a thin thermal plasma.
 
%[table2] modelfit
\begin{table}
\begin{center}
 \caption{Results of a thin thermal plasma model fit.}\label{tab:model}
 \begin{tabular}{lcc} \hline
  {Parameter} & \multicolumn{2}{c}{Value}\\ 
                 & {G355-a} & {G355-b} \\ \hline
    \(N_{\rm H} \) \ (\(\times\)10\( ^{22} \) cm\( ^{-2}\)) &    
   $5.8^{+1.0}_{-0.8}$  & $5.8^{+0.8}_{-0.7}$  \\ 
   $kT$\ (keV) & 
   $0.56^{+0.10}_{-0.10}$  & $0.56^{+0.08}_{-0.09}$ \\ 
   Si\footnotemark[$\ast$]\ & 
   $1.6^{+0.8}_{-0.5}$ & $1.6^{+0.6}_{-0.4}$ \\ 
   S\footnotemark[$\ast$]\ & 
   $3.1^{+0.9}_{-0.7}$   & $3.2^{+0.8}_{-0.7}$ \\ 
   Ar\footnotemark[$\ast$]\  & 
   $5.8^{+2.8}_{-2.1}$   & $5.8^{+2.4}_{-1.9}$  \\ 
   Ca\footnotemark[$\ast$]\  & 
   $15^{+27}_{-10}$   & $14^{+19}_{-8}$   \\ 
   Others\footnotemark[$\ast$]\  & 
   1.0\ (fixed) & 1.0\ (fixed) \\ 
 %  $\tau$\footnotemark[$\ddag$] ($\times10^{11}{\rm cm^{-3}s}$)  & ---   & ---  \\
   norm\footnotemark[$\dagger$]($\times10^{-2}$) & 
   $2.3^{+2.9}_{-1.2}$ & $2.3^{+2.4}_{-1.1}$ \\ 
   \(\chi^2\) /d.o.f. & 71.3/78 & 104.6/87 \\ \hline
  % Observed flux ($\times10^{-13}$ erg\ cm$^{-2}$\ s$^{-1}$)& 5.18  & 5.28 \\
  % $N_{\rm H}$-corrected flux ($\times10^{-11}$ erg\ cm$^{-2}$\ s$^{-1}$)& 1.59 & 1.23 \\ \hline
      \multicolumn{3}{@{}l@{}}{\hbox to 0pt{\parbox{80mm}{\footnotesize
      
	   \footnotemark[$*$] Abundance relative to the solar value \citep{solar}.
	    \par\noindent
	    \footnotemark[$\dagger$] Defined as $10^{-14}\times \int n_{\rm e}n_{\rm H}dV/(4\pi d^{2})$
	    , where {\it d} is the distance (cm), $n_{\rm H}$ is the hydrogen density ($\rm cm^{-3}$),
	     and {\it V} is the volume ($\rm cm^{3}$).
   }\hss}}
  \end{tabular}
 \end{center}
 \end{table}

%%%%%%%%%%%%%%%%%%%%%%%%%%%%%%%%%%%%%%%%%%%
We then applied a thin thermal plasma model in collisional ionization equilibrium (CIE) 
({\tt vapec} in XSPEC).
The abundance tables were taken from \citet{solar}.
First, the abundances were fixed to the solar values.
The model was rejected with the large $\Delta \chi^2$/d.o.f of 142/82 and 198/91 for G355-a and G355-b, respectively. 
Also we found positive residuals at the energies corresponding to the K-shell lines from He-like Si, S, Ar, and Ca.
Next, abundances of Si, S, Ar, and Ca were set to be free parameters,
but the others were fixed to be solar abundances.
This model well represents both G355-a and G355-b.
The best-fit parameters are listed in table \ref{tab:model}, while the best-fit model is plotted in figure 3.
We found that the abundances of Si, S, Ar, and Ca are larger than the solar values.

\subsection{Suzaku J173530$-$3236}

We analyzed spectra of the point source Suzaku J173530$-$3236 (=XMM J173530.9$-$323558), 
located at the northeast of G355.6$-$0.0.
We extracted the spectra from a circular region with a radius of 58$''$. 
The background subtraction was  the same as G355.6$-$0.0.
We made the near-sky-background-subtracted (hereafter Suzaku J1735-a) 
and the simulated-background-subtracted spectra (hereafter Suzaku J1735-b).
ARFs were created assuming a point-like source.

Figure 4 shows the background-subtracted spectra.
First, we applied a bremsstrahlung model modified by low-energy absorption to the spectra and found an emission line feature at $\sim$6.7 keV in the residuals.
The fit improved by adding an emission line model with a null line width:
$\Delta \chi^2$ values of 20.5/19.8 for Suzaku J1735-a/Suzaku J1735-b
indicate that the additional emission line is statistically significant 
at the $>99\%$ confidence level.
The best-fit parameters are listed in table 3, while the best-fit model is plotted in figure 4.  
The emission line at 6.72 keV is strong with an equivalent width of $\sim$1 keV.
The observed and the $N_{\rm H}$-corrected energy flux  in the 2--10 keV band is estimated to be $3.1\times 10^{-13}\ {\rm erg\ cm^{-2}\ s^{-1}}$ and $3.7\times 10^{-13}\ {\rm erg\ cm^{-2}\ s^{-1}}$, respectively.
	
\begin{figure*}
   \begin{center}
      \FigureFile(160mm,80mm){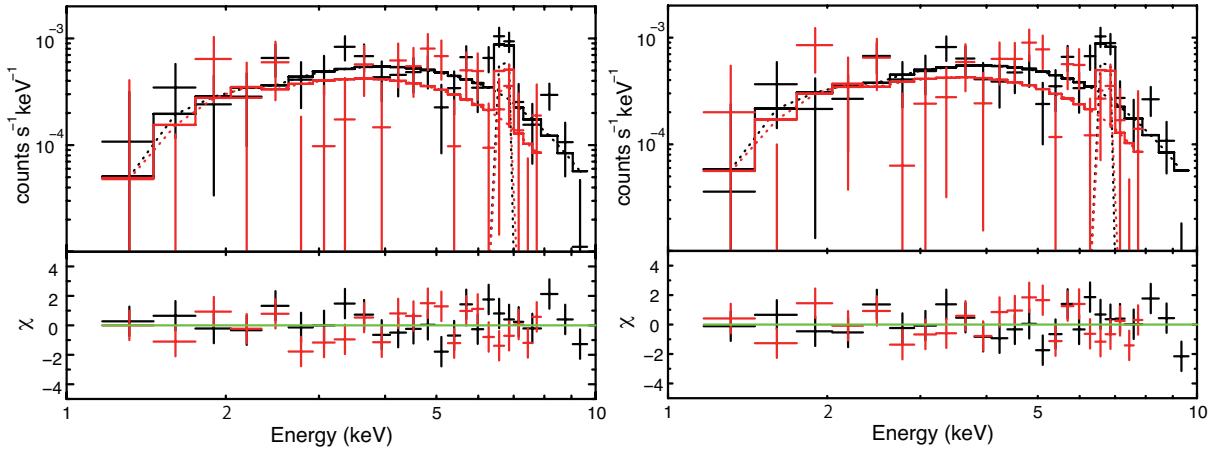}
%      \FigureFile(80mm,80mm){ronbun_j173530_b525.eps}
   \end{center}
   \caption{XIS spectra of Suzaku J173530$-$3236 and the best-fit model (upper), and 
   residuals from the best-fit model (lower). 
  The left panel is the results of Suzaku J1735-a (the near-sky-background-subtracted spectra), 
   while the right panel is those of Suzaku J1735-b (the simulated-background-subtracted spectra).
   Data of XIS0+3 and XIS1 are indicated in black and red, respectively.}\label{sp_point}
\end{figure*}

%[table3] modelfit
\begin{table}
\begin{center}
\caption{Results of a model fit for Suzaku J173530$-$3236.}\label{tab:point}
\begin{tabular}{lcc} \hline
Parameter      & \multicolumn{2}{c}{Value\footnotemark[$\ast$] } \\
                      &    Suzaku J1735-a & Suzaku J1735-b\\
 \hline
   $N_{\rm H}$\ (\(\times\)10\( ^{22} \)cm\( ^{-2}\)) & $3.4^{+1.7}_{-1.1}$ & $3.2^{+1.5}_{-1.0}$\\ 
   {\it kT}\ (keV) & $>$12.9 & $>$10.3\\ 
   {\it E}\ (keV) & $6.72^{+0.05}_{-0.04}$ & $6.72^{+0.04}_{-0.05}$\\ 
   {\it EW}\footnotemark[$\dagger$]\ (keV) & 1.20$\pm$0.27 & 1.17$\pm$0.27\\ 
   \(\chi^2\) /d.o.f. & 45.2/43 & 49.6/43\\
 \hline
 \multicolumn{2}{@{}l@{}}{\hbox to 0pt{\parbox{80mm}{\footnotesize
	   \footnotemark[$\ast$] A model consisting of bremsstrahlung and a Gaussian line.
	    \par\noindent
	    \footnotemark[$\dagger$] An equivalent width.
	        }\hss}}
   \end{tabular}
\end{center}
\end{table}

%%%%%%%%%%%%%%%%%%%%%%%%%%%%%%%%%%%%%%%%%%%%%%%%%%	
\section{Discussion}

The excellent capability of the Suzaku XIS enable us to resolve the diffuse emission detected with ASCA 
at the position of G355.6$-$0.0 into two objects, 
G355.6$-$0.0 and Suzaku J173530$-$3236.
The X-ray spectrum of G355.6$-$0.0 is due to a thin thermal plasma of $\sim 0.6$ keV temperature,  
while that of Suzaku  J173530$-$3236 is hard
with a strong iron emission line.
We confirmed that the Suzaku best-fit models of G355.6$-$0.0 and Suzaku J173530$-$3236 well 
reproduce the ASCA spectrum. 
This indicates that both G355.6$-$0.0 and Suzaku J173530$-$3236 show 
no obvious flux and spectral change.

We used the X-ray spectra after two different background subtraction, 
the near-sky-background-subtracted (G355-a and Suzaku J1735-a) and 
the simulated-background-subtracted spectra (G355-b and Suzaku J1735-b). 
The results obtained from the two methods are essentially the same.
We discuss the characteristics of G355.6$-$0.0 and Suzaku J173530$-$3236
based on the results of the spectra with better statistics (G355-b and Suzaku J1735-b).
	
\subsection{SNR G355.6$-$0.0}

%%%Distance to G355.6$-$0.0

The column density of the hydrogen atom ($N_{\rm H_{I}}$) along the line of sight to G355.6$-$0.0 
is $\sim$1.7$\times$$10^{22}$ cm$^{-2}$ \citep{H1} 
and that of the hydrogen molecule ($N_{\rm H_{2}}$) is (2.3--3.5)$\times$$10^{22}$ cm$^{-2}$, 
where the conversion factor (X-factor) from CO to H$_2$ is assumed to be 
1.8$\times$10$^{20}$ cm$^{-2}$ K $^{-1}$ km $^{-1}$ s \citep{H2}. 
Then, using the mean value of  $N_{\rm H_{2}}$=2.9$\times$$10^{22}$ cm$^{-2}$, 
the total column density can be estimated to be 
$N_{\rm H}=N_{\rm H_{I}}+2N_{\rm H_{2}}\sim$7.5$\times10^{22}{\rm cm}^{-2}$.
Since the observed $N_{\rm H}$ value (5.8$\times10^{22}{\rm cm}^{-2}$) is 77\% of this value, 
the distance to G355.6$-$0.0 can be 77\% of 17 kpc (double the distance from the Sun to the Galactic center), which is 13 kpc.
The distance estimated by the best-fit $N_{\rm H}$ value is consistent with the distance of 12.6 kpc, derived from the 
empirical $\Sigma$-D relation in radio \citep{sigma-D}.
We, therefore, adopted the distance to be 13 kpc in the following discussion.
The X-ray luminosity is calculated to be 3$\times10^{35}\ \rm{erg\ s^{-1}}$ in the 1--10 keV band.

%%%Properties of G355.6$-$0.0

Since the spatial resolution of Suzaku is limited, we cannot firmly conclude
from the Suzaku data alone, 
whether the X-ray emission is center-filled or shell-like. 
We, therefore, consulted the XMM data in the observation of NGC6383. 
%%%%%%%%%%%%%%%%%%%%%%%%%%%%%%%%%%%%%%%%%%%%
From the XMM image in the 1--5 keV band, 
we found faint and extended  X-rays at the position of G355.6$-$0.0, 
which is not shell-like but center-filled.  
Assuming the plasma is a sphere with a radius of $r_{\rm x}$=7.6 pc 
(the size of  2$'$ at the distance of 13 kpc), 
X-ray emitting volume 
$V=(4/3) \pi r_{\rm x}^{3}f$
 is estimated to be 
$5.5\times 10^{58}f\ \rm cm^3$, where {\it f} is a filling factor.
The best-fit emission integral is ${\it E.I.}=\int n_{\rm H}n_{\rm e}dV\sim n_{\rm H}n_{\rm e}V$ = 
$4.7\times 10^{58}\ \rm cm^{-3}$, where $n_{\rm H}$ and $n_{\rm e}$ are the hydrogen and 
electron densities, respectively.
Then with the assumption of $n_{\rm e}=1.2n_{\rm H}$, the hydrogen density and the total gas mass ({\it M}), 
are calculated to be $n_{\mathrm{H}}=0.85 f^{-0.5}\ \rm cm^{-3}$, $M\sim1.4 n_{\rm H}m_{\rm H}V
\sim55M_{\odot}f^{0.5}$, respectively, where $m_{\rm H}$ is the hydrogen atomic mass.
The plasma must be expanding with the sound velocity of 
$v_{\rm s}=(
%\frac{\gamma kT}{\mu m_{\rm H}}
\gamma kT/\mu m_{\rm H}
)^{0.5}$, where $\gamma$=5/3 and $\mu$=0.60, 
then the dynamical age, $t$, is estimated to be $t$=$r_{\rm x}$/$v_{\rm s}\sim2\times10^4$ yr.
This means that G355.6$-$0.0 is a middle-aged or old SNR.

%%% MM-SNRs and RP

As we noted, G355.6$-$0.0 has a center-filled X-ray plasma within the radio shell.
Therefore, G355.6$-$0.0 is considered to be a member of MM SNRs.
The thermal plasmas of most of the  MM SNRs are typically in the CIE state with the metal abundances 
comparable to the solar values \citep{mm}.
Thus, the plasma has been considered to be interstellar origin, not ejecta origin. 
However, since G355.6$-$0.0 exhibits a CIE spectrum with enhanced abundances, 
the plasma prefers the ejecta origin. 
%\textcolor{red}{The abundance pattern is similar to that of the type Ia SN rather than core-collapsed SN}
%\textcolor{red}{
%(e.g., \cite{Tsujimoto1995, Iwamoto1999, Kobayashi2006}).}

Recently, an over-ionized plasma was discovered from some MM SNRs 
\citep{rrc1,rrc2,rrc3,rrc4,rrc5,rrc6}. 
Although their ages are not well determined, all may be middle-aged SNRs. 
Since the electron temperatures of MM SNRs with the over-ionized plasma are in the range of 0.3--1.5 keV, 
G355.6$-$0.0 can be also a candidate of the over-ionized plasma emitter. 
%%%%%%%%%%%%%%%%%%%%%%%%%%%%%
The G355.6$-$0.0 spectrum, however,  shows no clear sign of over-ionization 
such as the negative residual in the 2--3 keV band as is found in G359.1$-$0.5 and G346.6$-$0.2 \citep{rrc3,rrc6}. 
Furthermore, the recombination timescale, which is derived from spectral analysis with a non-equilibrium ionization plasma model, is large enough to become CIE.
What makes this difference?
The MM SNRs having a over-ionized plasma share some common characteristics; all are accompanied with OH masers and most of them are either GeV or TeV $\gamma$-ray sources \citep{rrc1,rrc2,rrc3,rrc4,rrc5,rrc6}.
These suggest that the SNRs interact with the surrounding molecular clouds.
However, neither OH masers nor $\gamma$-ray emission are found in G355.6$-$0.0.
These differences might be related to the structures of the surrounding environments.
To address this question, further  systematic study of MM SNRs is necessary.
 
\subsection{Hard X-ray Source Suzaku J173530$-$3236} 

Since the $N_{\mathrm{H}}$ value of Suzaku J173530$-$3236 is smaller than that of G355.6$-$0.0, 
Suzaku J173530$-$3236 is located at near side of G355.6$-$0.0.
%The observed $N_{\rm H}$ value of 3.2$\times10^{22}$ cm$^{-2}$ is 43\% of 7.5$\times10^{22}$ cm$^{-2}$, and hence the distance would be 7.3 kpc.
The observed $N_{\rm H}$ value of 3.2$\times10^{22}$ cm$^{-2}$ corresponds to 43\% of 7.5$\times10^{22}$ cm$^{-2}$, yielding the distance of 7.3 kpc.

The spectral features, a thin thermal spectrum with the emission line at $\sim$ 6.7 keV, 
are similar to those observed in cataclysmic variables (CVs).
Moreover, the luminosity of 2$\times10^{33}$ erg s$^{-1}$ in the 2--10 keV band is also 
in the range of CVs (e.g., \cite{CV1}, \cite{CV2}).
Thus, Suzaku J173530$-$3236 is likely to be a CV-like star.
We note that the equivalent width of the Fe line is larger than typical values of CVs 
\citep{CV1,CV2}.
Due to the limited photon statistics, however, the nature of 
Suzaku J173530$-$3236 is inconclusive by  the X-ray data alone.
Follow up observations in the near-infrared band are required.\\
	
%acknow ledgement
The authors are grateful to all the members of the Suzaku team.
This research made use of the NASA/IPAC Extragalactic Database (NED) 
operated by Jet Propulsion Laboratory, California Institute of Technology, 
under contract with NASA and 
the SIMBAD database operated at the CDS, Strasbourg, France. 
This work was supported by
the Japan Society for the Promotion of Science (JSPS); the Grant-in-Aid 
for Scientific Research (C) 21540234 (SY), 24540232 (SY), and 24540229 (KK), 
Challenging Exploratory Research program  20654019 (KK), 
and Specially Promoted Research 23000004 (KK).

%%%%%%%%%%%%%%%%%%%%%%%%%%%%%%%%%%%%%%%
%%%
% See the manual for the detail.
%%%

\end{document}